\documentstyle{l-aa}

\def\beq{\begin{equation}}
\def\eeq{\end{equation}}
\def\bc{\begin{center}}
\def\ec{\end{center}}

   \thesaurus{12.12.1; 13.07.1}

   \title{Anisotropy of the sky distribution of gamma-ray bursts}

   \author{L.G. Bal\'azs$^1$, A. M\'esz\'aros$^{1,2,3}$, I. Horv\'ath$^4$}

\offprints{L.G. Bal\'azs}

\institute{$^1$Konkoly Observatory, Budapest, Box 67, H-1505 Hungary;
e-mail: balazs@ogyalla.konkoly.hu \\
$^2$Department of Astronomy, Charles University, 180 00 Prague 8,
V Hole\v{s}ovi\v{c}k\'ach 2, Czech Republic; meszaros@mbox.cesnet.cz\\
$^3$European Southern Observatory, Garching bei M\"{u}nchen,
K. Schwarzschild-Str. 2, Germany\\
$^4$Department of Earth Science, Pusan National University, Pusan,
609-735, Korea (South); e-mail: hoi@astrophys.es.pusan.ac.kr
}
   \date{Received March ..., 1998 / accepted ...}

\begin{document}

   \maketitle

\begin{abstract}

The isotropy of gamma-ray bursts collected in Current BATSE catalog is
studied. It is shown that the
quadrupole term being proportional to $\sim \sin 2b \sin l$ is non-zero
with a probability 99.9\%. The occurrence of this anisotropy term is then
confirmed by the binomial test even with the probability 99.97 \%. Hence,
the sky distribution of all known gamma-ray bursts
is anisotropic. It is also argued that this anisotropy cannot be caused
exclusively by instrumental effects due to the nonuniform sky exposure
of BATSE instrument.
Separating the GRBs into short and long subclasses, it is shown that
the short ones are distributed anisotropically, but the long ones seem
to be distributed still isotropically. The character of anisotropy
suggests that the cosmological origin
of short GRBs further holds, and there is no evidence for their
Galactical origin.


\end{abstract}

\section{Introduction}

The true physical nature of gamma-ray bursts (GRBs) is one of the tantalizing
enigmas of the recent astrophysics.
 Although since their first detection (Klebesadel et al. 1973) there
were several
suggestions trying to give a clear explanation for their origin, no definite
answer has been given yet (cf. Paczy\'nski 1995).
 Recently, the successful identifications made
by the Beppo-SAX satellite, followed by the detection of optical counterparts
(van Paradijs et al. 1997),
seem to give a firm support for the models aiming to explain the bursts by
merging neutron stars (Usov \& Chibisov 1975;
Rees \& M\'esz\'aros P. 1994; M\'esz\'aros P. \& Rees 1997) and seem to
put them definitely into the cosmological distances. The alternative
Galactical origin seems to be ruled out (a survey of the question of
distances may be found, e.g., in Paczy\'nski 1995). However,
the small number of optically identified events is far from being enough to
characterize the properties of the whole burst population. On the other
hand, the existence of cosmological distances of GRBs seems to be definite.

In addition, even before this identification, indirect
observational evidences were known for the cosmological origin. These
evidences were based mainly on the modifications of $<V/V_{max}>$ test
(cf. Norris et al. 1994; M\'esz\'aros P. \& M\'esz\'aros A. 1995;
Norris et al. 1995; Nemiroff 1995; Horv\'ath et al. 1996), and on the
study of the time dilatation (cf. Norris et al. 1995;
M\'esz\'aros A. \& M\'esz\'aros P. 1996; Stern 1996; M\'esz\'aros A.
et al. 1996;
 Che et al. 1997). A further important indirect support of
cosmological origin is based on the observed isotropy on the sky
(Briggs 1993; Syer \& Saha 1994; Briggs 1995;
Tegmark et al. 1996a,b; Briggs et al. 1996).
All these papers suggest that the angular distribution is
isotropic, because there are no statistically significant departures from
the isotropy. Also the separations of GRBs into the different subclasses
either due to the duration (Kouveliotou et al. 1993; Belli 1995;
 Dezalay et al. 1996) or
due to the fluence on channel above the 300 keV
(Pendleton et al. 1997) do
not change the situation; the proposed subclasses alone also seem
to be distributed isotropically. Hence, it can well occur that
the total number of observed GRBs is a mixture of a wide variety of
physically different objects, but all GRBs should be at cosmological
distances due to their isotropic angular distribution on sky, and due
to other direct and indirect supports.

In this paper we find a clear anisotropy of all GRBs, and then
separately of the short GRBs, too. On the other hand, the long GRBs
seem to be still distributed isotropically.
To do this we will use both the standard analysis based on the
spherical harmonics and also the so called binomial test.
It may seem that all this is an argument against the cosmological origin
of short GRBs. It is shown that this is not the case.

The paper is organized as follows. In Section 2 the key ideas of
the used statistical tests
are recapitulated. In Section 3 the anisotropy of all GRBs
is demonstrated. In Section 4 the anisotropy of short and the isotropy of long
GRBs is shown. Section 5 discusses the results, and argues for the cosmological
origin of short GRBs. Finally, in Section 6
there is a summation of results.

\section{Mathematical skeleton of the problem}

Testing the isotropy on the celestial sphere one may use several methods.
Nevertheless, strictly from the mathematical point of view,
the necessary condition for the isotropy is the
stochastic independency of the sky distribution of the bursts on their
observed physical properties.  It means that, if
$f(b,l, x_1,...,x_n)dF dx_1...dx_2$ is the
probability of finding an object in the $dF=\cos b\; dl\;db$ infinitesimal
solid angle and
in the $(x_1, x_1+dx_1, ..., x_n,x_n+dx_n)$ interval, one must have
\begin{equation}
f(l,b,x_1,...,x_n) = \omega(l,b) g(x_1,...,x_n).
\end{equation}
Here
$0\leq l \leq 360^o,\; -90^o \leq b \leq 90^o$ give the celestial positions
in Galactical coordinates,
$x_n$ ($n \geq 1$) measure the
physical properties (peak fluxes, fluences, durations, etc...) of GRBs
and $g$ is their probability density.
It may well be assumed that in the case of isotropy the distribution of GRBs
fulfils this equation (cf. Briggs et al. 1996; Tegmark et al. 1996a,b).

However, statement (1) is only a necessary but not a
sufficient condition for isotropy.
Isotropy means that also $\omega(l,b)=1/(4\pi) $. Hence,
in this case, for $N$ observed GRBs the events $dN=N \omega dF$, i.e.
the expected number of GRBs in an infinitesimal solid angle, is
not depending on $[l,b]$. In other words, the isotropy
means that the probability of observation of a
burst in a solid angle $0 \leq \Omega
\leq 4 \pi$ ($\Omega$ is in steradians) is given by
$\Omega/(4 \pi)$ and is independent on its location on the celestial sphere.
This follows immediately from (1), if one does integration over $l$
and $b$ to obtain, first, the solid angle $\Omega$, and, second, the
whole sky. Then the ratio of two results gives $\Omega/(4 \pi)$,
and the concrete form of $g$ is unimportant.

The most frequently used procedure to test the isotropy of GRBs
is based on the spherical harmonics (Briggs 1993, 1995;
Briggs et al. 1996;
Tegmark et al. 1996a,b). The key idea is the following.
In general case one may decompose the function $\omega(b,l)$ into the
well-known spherical harmonics. One has:
$$
\omega(b,l) = (4\pi)^{-1/2} \omega_0 - $$
$$(3/(4\pi))^{1/2} (\omega_{1,-1} \cos b \sin l
- \omega_{1,1} \cos b \cos l
+  \omega_{1,0} \sin b) +$$
$$ (15/(16\pi))^{1/2}( \omega_{2,-2} \cos^2 b \sin 2l
+  \omega_{2,2} \cos^2 b \cos 2l - $$
$$   \omega_{2,-1} \sin 2b \sin l
-   \omega_{2,1} \sin 2b \cos l) + $$
\begin{equation}
(5/(16\pi))^{1/2}\omega_{2,0} (3 \sin^2 b - 1)  +  higher\, order\,
 harmonics.
\end{equation}

The first term on the right-hand side is the monopole term, the
following three ones are the dipole terms, the following five ones are the
quadrupole terms
(cf. Press et al. 1992; Chapt. 6.8). Nevertheless,
$\omega$ is constant for isotropic distribution, and hence on the right-hand
side any terms,
except for $\omega_0$, should be identically zeros. To test this hypothesis
one may proceed, e.g., as follows. Let there are observed
$N$ GRBs with their
measured positions $[b_j, l_j]$ ($ j = 1,2,...,N$). In this case $\omega$
is given as a set of points on the celestial sphere.
Because the spherical harmonics are orthogonal functions, to calculate the
$\omega_{\{\}}$ coefficients one has to compute the functional scalar products.
For example, $\omega_{2,-1}$ is given by
$$
\omega_{2,-1}= - \Bigl(\frac{15}{16\pi}\Bigr)^{1/2}
\int \limits_{-\pi/2}^{\pi/2} \cos b \; db
\int \limits_{0}^{2 \pi} \;
\omega(l,b) \sin 2b \sin l\;dl $$
\begin{equation}
= - (15/(16\pi))^{1/2}N^{-1}\sum \limits_{j=1}^N
 \sin 2b_j \sin l_j.
\end{equation}
Because $\omega$ is given only in discrete points, the integral is transformed
into an ordinary summation (cf. Kendall \& Stuart 1969, p. 16).
In the case of isotropy one has $\omega_{2,-1}=0$,
and hence  $N^{-1}\sum \limits_{j=1}^N \sin 2b_j \sin l_j = 0$.
Therefore,
the expected mean of $\sin 2b_j \sin l_j$ values is zero. One has to
proceed similarly to any other $\omega_{\{\}}$ coefficient.

In order to test the zero value of, e.g., $\omega_{2,-1}$ one has to calculate,
first,
$\sin 2b_j \sin l_j$ for any $j = 1,2,...,N$,
and, second, the mean, standard deviation and Student's $t$ variable
(cf. Press et al. 1992, Chapt. 14). Finally, third, one has to ensure
the validity of zero mean
from Student test. As far as it is known, no statistically
significant anisotropies of GRBs were detected yet by this procedure
(cf. Briggs et al. 1996; Tegmark et al. 1996a,b).

Nevertheless, there are also other ways to test the isotropy.
An extremely simple method uses the binomial distribution.
In the remaining part of this section we explain this test (see also
M\'esz\'aros A. 1997).

In order to test the anisotropy by this method
one may proceed as follows. Let us given an area on the sky defined
by a solid angle $0 < \Omega < 4 \pi$ ($\Omega$ is in steradians).
In the case of isotropy the $p$ probability
to observe a burst within this area is $p = \Omega/(4\pi)$. Then, obviously,
$q=1-p$ is the probability to have it outside.
  Observing $N > 0$ bursts on the whole sky the
 probability to have $k$ bursts (it may be
 $k=0,1,2,...,N$) within $\Omega$ is given
 by the binomial (Bernoulli) distribution taking the form
\begin{equation}
P_p(N,k)= {N!\over k!(N-k)!}p^k q^{N-k}.
\end{equation}
This distribution is one of the standard probability distributions
discussed widely in statistical
text-books (e.g. Trumpler \& Weaver 1953; Kendall \& Stuart 1969, p.120;
about its use in astronomy see, e.g., M\'esz\'aros
1997). The expected mean is $Np$ and the expected variance is
 $Npq$.
One may also calculate the integral (full) probability, too, by
a simple summation.

In our case we will consider $N$ GRBs, and we will test the hypothesis
whether they are distributed isotropically on the sky.
Assume that $k_{obs}$ is the
observed number of GRBs at the solid angle $\Omega$.
If the apriori assumption is the isotropy, i.e $p = \Omega/(4\pi)$,
then one may test whether the observed number $k_{obs}$ is compatible with
this apriori assumption. Of course, any $0 \leq k_{obs} \leq N$ can occur
with a certain probability given by the binomial distribution.
But, if this probability is too small, one
hesitates seriously to accept the apriori assumption.

Consider the
value $\mid k_{obs.}-Np \mid = k_o$. The value $k_o$
characterizes "the departure" of $k_{obs}$ from the mean
$Np$. Then one may introduce the probability
$$
P(N, k_{obs}) = \;\;\;\;\;\;\;\;\;\;\;\;\;\;\;\;\;\;\;\;\;\;\;\;\;\;\;\;\;$$
\begin{equation}
 1 - P_{p,tot}(N,(Np +k_o)) +
 P_{p,tot}(N,(Np - k_o)).
 \end{equation}
$P(N,k_{obs})$ is the probability that
the departure of $k_{obs}$ from the $Np$ mean is still given by a chance.

In order to test the isotropy of the GRBs celestial distribution
we will divide the sky into two equal areas, i.e we will choose $p =
0.5$.
It is essential to note here that neither of these two parts
must be simply connected compact regions.

\section{Anisotropy of all GRBs}

 In order to test the isotropy of GRBs we will test the three dipole
and five quadrupole terms in accordance with the method described in
the previous Section.
We will consider all GRBs that have well-defined angular positions. Up
to the end of year 1997 there were 2025 such objects at Current BATSE catalog
(Meegan et al. 1997; Paciesas et al. 1998).
The results are summarized in Table 1.

   \begin{table}
     \label{Student1}
$$
         \begin{array}{lcr}
            \hline
            \noalign{\smallskip}
  & t   &  \% \cr
            \noalign{\smallskip}
            \hline
            \noalign{\smallskip}
   \omega_{1,-1}& 1.51   & 13.4    \cr
   \omega_{1,1}& 1.77  &  7.7  \cr
   \omega_{1,0}& 0.71  &  47.7   \cr
   \omega_{2,-2}& 2.76  & 0.6   \cr
   \omega_{2,2}&  1.54  & 12.1    \cr
   \omega_{2,-1}& 3.26  & 0.1     \cr
   \omega_{2,1}& 0.98  &  33.3   \cr
   \omega_{2,0}& 0.36  &  71.9   \cr
            \hline
         \end{array}
$$
      \caption{Student test of the dipole and quadrupole terms of 2025
GRBs. In the first column the coefficients defined in Eq.2 are given.
In the second column the Student $t$ is provided. The third column
shows the probability that the considered terms are still zeros.}

\end{table}

We see that, except for the terms defined by $\omega_{2,-1}$ and
$\omega_{2,-2}$, the remaining six terms may still be taken to be zero.
This means that {\it there is} a clear anisotropy defined by term
$\sim \sin 2b \sin l$.  The probability that this term is zero is smaller
than $0.27 \%$. It is practically sure that the second quadrupole
term being proportional to $\sim \cos^2 b \sin 2l $ is non-zero, too.
Nevertheless, we will not deal with
this second term in this paper, because the purpose of this paper
is to demonstrate only {\it qualitatively}
the existence of anisotropies.

The anisotropy defined by $\omega_{2,-2}$ may be defined
in another form, too. This quadrupole term has a positive sign, when both $b$
and $l$ have the same signs, and has a negative sign, when $l$ and
$b$ have opposite signs. Therefore, let us define two parts of the sky
having the same sizes ($2 \pi$ steradians). The first one is defined by
Galactical coordinates $b > 0,\; l > 180^o$ and $b < 0,\; 0< l < 180^o$. This
means that this first
part is in fact composed from two separated "sky-quarters". The second part
is then given by $b > 0,\; 0 < l < 180^o$ and $b < 0, \; l > 180^o$. This
means that this second part is also given by two separated
"sky-quarters". Then the detected quadrupole anisotropy suggests that there
should be an essential difference, e.g., in the number of GRBs in these
two parts.

In order to test again this expectation we will do
the test based on Bernoulli distribution. We divide
the whole sky into these two parts, and hence we expect a
Bernoulli distribution with $p=0.5$ for $N= 2025$.  As it is noted
in Section 2, it is certainly allowable that these parts are composed from
several subregions.

 A straightforward counting of GRBs in these regions shows
that 930 GRBs are in the first one and 1095 are in the second. (Note that
no GRBs had coordinates exactly either $b= 0$, or $b = \pm 90^o$, or
$l= 0$, or $l = 180^o$. This means that in the paper
no problems have arisen from the fact that the
boundaries of "sky-quarters" were not taken into acount.)
Assuming $p=0.5$ the binomial (Bernoulli) test gives a $0.03\%$
probability that this distribution
is caused only by a chance. Hence, the relatively smaller number in the
first region compared with the second one is {\it not} a chance,
and the distribution of all GRBs is anisotropic with a certainty.

Clearly, concerning the consequences of the intrinsic
anisotropy of GRBs, one must be still careful.
Instrumental effects of the BATSE experiment may also play a role, and in
principle
it can also occur that the detected anisotropy is caused exclusively by
instrumental effects. To be as correct as possible, one may claim that,
 in essence, there can be three
different causes of this observed anisotropy: a. The anisotropy is purely
caused by the BATSE's nonuniform sky exposure (in other words, the
intrinsic angular distribution of GRBs is still isotropic, and the observed
anisotropy is a pure instrumental effect); b. The anisotropy is purely
given by the intrinsic anisotropy of GRBs, and the instrumental effects are
unimportant; c. The anisotropy is given both by instrumental effects and
also by the intrinsic anisotropy. To be sure that there is also an
intrinsic anisotropy of GRBs, one must be sure that the possibility a. does
not occur. In what follows, when we will speak about "the possibility a.",
we will consider this one.

   It is well-known that the sky exposure of the BATSE instrument is
nonuniform. This question is described and discussed in several papers
(cf. Briggs 1993; Fishman et al. 1994; Tegmark et al. 1996a,b; Briggs et al.
1996). The BATSE sky coverage depends on the declination only
in the equatorial
coordinate system (Tegmark et al. 1996b) in
such a manner that the probability of
detection is about 10\% higher near the pole than near the equator. This
behaviour in Galactic coordinates predicts excess numbers of GRBs just in
sky-quarters given by $b > 0,\; 0 < l < 180^o$ and $b < 0, \; l > 180^o$.
Hence, in principle, it is well possible that the observed
anisotropy is caused by a pure instrumental effect. The purpose of
Section 4 is to show that this is {\it not} the case, and the possibility a.
should be excluded.

\section{Different distribution of short and long GRBs}

If there is an intrinsic isotropy indeed, then
Eq.(1) will be further fulfilled, and
$\omega(b,l)$ itself will reflects the nonuniformity of sky-coverage.
Then at the first part the number of observed GRBs
should be smaller, because the integrations of
$\omega (l,b)dF$ giving the first and second part, respectively, do not
give the same values. Their ratio should be $\simeq 930/1095 = 0.85$.
This ratio should be obtained for any subset of GRBs, when the choice
of this subset is based on some
physical properties of the bursts, because the function $g$ does not enter
into the calculation for $X$. In other words,
if the level of anisotropy depends on the
physical parameters of GRBs, then "the possibility a." should be excluded.

To be extra cautious it is also necessary to remember that some bias
may arise for the dimmest GRBs, because for them it is not necessarily
true that the total exposure time is exactly proportional to the
observed number of sources due to the varying threshold limit of BATSE
(see Tegmark et al. 1996a for the discussion of this question).
To avoid this bias the simplest procedure is not to take into
account the dimmest GRBs.

 All this allows to exclude the possibility a. quite simply.
To do this it is necessary to take some subsets of GRBs, and to verify
for any of them that the ratio X is roughly 0.85. For security, it is
also necessary to omit the dimmest GRBs.

For the sake of maximal correctness we will not use any ad hoc criterions
to define such subsets, but we will exclusively use criterions which were
introduced earlier by others. First, we exclude any GRBs having the peak fluxes
on 256 ms trigger smaller than 0.65 photons/(cm$^2$s). The truncation
with this threshold is proposed and used in Pendleton et al. (1997).
Second, from the remaining
GRBs we exclude GRBs which have no defined duration $T_{90}$
(for the definition of this duration see Kouveliotou et al. 1993).
Third, we also exclude the bursts which have no $f3$ value, which is the
fluence on the energy channel [100, 300] keV. These truncations are also
necessary, because we will consider the subsets,
for which the criterions use $T_{90}$ and $f3$.

932 GRBs remain, and from them 430 are at the first part, and 502 are at
the second one. Hence, here X = 430/502 = 0.86. There is no doubt
that this "truncated" sample of GRBs is distributed similarly to that of
whole sample with 2025 GRBs. The probability that this distribution is given
by a chance is here 2\%.
(Note that, of course, here the same X does not give the
same probability, because there is a smaller number in the sample.)

 932 GRBs will be separated, first, into the "short" and "long"
subclasses (cf. Kouveliotou et al. 1993), and, second, into the
No-High-Energy (NHE) and High-Energy (HE) bursts (Pendleton et al.
1997).

    The boundary between the short and long bursts is usually taken
for $T_{90} = 2$ s (Kouveliotou et al. 1993; Belli 1995;
Dezalay et al. 1997). Nevertheless, this
 boundary at $T_{90} = 2$ s is not so precise
(e.g. in Katz \& Canel (1996) $T_{90} = 10$ s is used). In addition,
this boundary gives no definite strict separation, because
at class $T_{90}< 2$ s
long bursts, and at class $T_{90} > 2$ s short bursts are also possible,
respectively (cf. Belli 1995).
Therefore, in order to test more safely the
distribution of two subclasses, we will consider also the case
when the boundary is at
$T_{90} = 10$ s. We will consider also the subsamples of GRBs
having $T_{90} < 1$ s, and $T_{90} > 15$ s, respectively, because
then they contain surely only short and long bursts, respectively. The results
are shown in Table 2.

   \begin{table}
     \label{Short-long}
$$
         \begin{array}{ccrcc}
            \hline
            \noalign{\smallskip}
sample & N & k_{obs} & (N-k_{obs}) & \% \\
            \noalign{\smallskip}
            \hline
            \noalign{\smallskip}
all \phantom{@}GRBs & 932 & 430 & 502 &  2.0 \\
T_{90}<1s & 206  &  82   & 124 & 0.43 \\
T_{90}<2s & 251  & 103 & 148  &  0.55  \\
T_{90}<10s & 372  & 154 & 218  &  0.11  \\
T_{90}>2s & 681  & 327 & 354  &  32\phantom{@}    \\
T_{90}>10s & 560  & 276 & 284  &  77\phantom{@}    \\
T_{90}>15s & 507  & 247 & 260 & 59 \phantom{@}  \\
\hline
\end{array}
$$
\caption{Results of the binomial test of subsamples of GRBs with different
durations. $N$ is the number of GRBs at the given subsample,
$ k_{obs}$ is the observed number GRBs at the first part in this subsample,
  and \% is the probability in
percentages that the assumption of isotropy is still valid.}

\end{table}

Table 2 shows, e.g., that there are 251 GRBs with
$T_{90} < 2$ s, and 681 GRBs with $T_{90} > 2$ s.
Then, from the short GRBs 103 are at the first part of sky
and 148 at the second one.
This gives X = 103/148 = 0.70. It seems that
at the first part there is even a smaller portion of shorter GRBs
than that of the all GRBs.
The probability that this is a chance is given by 0.55\%.
 On the other hand,
from the long GRBs 327 are at the first part and 354 at the second one.
This gives X = 327/354 = 0.92. Hence, it seems immediately that for the long
GRBs the isotropy is still an acceptable assumption.
The binomial test quantifies:
there is a 32\% probability that this distribution is given by a
chance. Doubtlessly, the long GRBs are distributed more isotropically
than the short ones; there is no statistically significant departure
from isotropy for long GRBs. The subsamples $T_{90} < 1$ s and
$T_{90} > 15$ s confirm this expectation; the boundary at $T_{90} = 10$
s also does not change the conclusion.
One may claim that the anisotropy of short GRBs is
statistically significant, but for long GRBs it is not.

Doubtlessly, the short and long subclasses are distributed differently.
This also excludes the possibility a.;
the observed anisotropy of all GRBs cannot be caused exclusively by
instrumental effects. It is difficult to imagine an instrumental effect
which leads to isotropy of long GRBs and to anisotropy of short GRBs.

Pendleton et al. (1997) introduces the subclasses of HE and NHE bursts.
The criterion depends on the ratio $f4/f3$, where $f3$ is the fluence
on energy channel $[100,300]$ keV and $f4$ is the fluence on the energy
channel $> 300$ keV.
(From this it is also clear, why we needed non-zero $f3$.
On the other hand, $f4$ can be vanishing; these GRBs are simply NHE
bursts.) Application of this criterion is not so simple, because for a great
portion of GRBs there are large uncertainties of the values of $f4$
due to their errors.
Concretely, for 693 GRBs the value of $f4$ is bigger than the corresponding
error of this $f4$;
for 131 GRBs there is no $f4$; for the remaining 108 GRBs there
are some values of $f4$ at the Current BATSE catalog (Meegan et al. 1997),
but they are smaller than their errors. Hence, 693 GRBs can be taken as HE
bursts (HE1 subsample). 131 GRBs having no $f4$ may be taken as NHE bursts
(NHE1 subsample). We did binomial tests for
these two subclasses. Separation of the remaining 108 GRBs
into the HE and NHE subclasses is not so clear. We consider artificially
the boundary as follows: If the value of $f4$ is bigger than the half of
error, then we have HE; otherwise NHE. Applying this criterion we
will have 168
NHE (NHE2 subsample) and 764 HE bursts (HE2 subsample).
For them the binomial tests were also done.
The results are collected in Table 3.

   \begin{table}
     \label{NHE-HE}
$$
         \begin{array}{ccccr}
            \hline
            \noalign{\smallskip}
sample & N & k_{obs} & (N-k_{obs}) & \% \\
            \noalign{\smallskip}
            \hline
            \noalign{\smallskip}
all \phantom{@}GRBs & 932 & 430 & 502 &  2.0 \\
NHE1 & 131   &  52   & 69 & 14.9 \\
HE1 & 693  & 327 & 366  &  14.6  \\
NHE2 & 168  & 69 & 99  &  2.5    \\
HE2 & 764  & 361 & 403 & 13.8   \\
\hline
\end{array}
$$
\caption{Results of the binomial test of NHE-HE subsamples of GRBs.
The subsamples NHE1, HE1, NHE2, HE2 are explained in the text.}
\end{table}

Table 3. gives an ambiguous result. Due to the smaller number in subclasses
no anisotropy is confirmed yet on a satisfactorily high
level of significance. In addition, contrary to the short-long separation,
there is no obvious difference between HE and NHE classes.
Hence, there is no obvious and unambiguous result here.

\section{Discussion}

The quadrupole anisotropy reported in this paper is an unexpected
and new result. As far as it is known no
anisotropy terms were detected yet (cf. Tegmark et al. 1996a,b).
Probably this situation was given by the fact that the
majority of these isotropy studies concentrated the effort into
dipole and $\omega_{2,0}$ quadrupole terms, which are expected to differ
from zero, if the GRBs are arisen in the Galaxy.

The essentially different angular distribution of short and long GRBs
suggest that their separation into these two subclasses has a deeper
cause. It is well-known that in average the short bursts have higher
hardnesses (hardness = $f3/f2$, where $f2$ is the fluence on energy channel
[50, 100] keV). Katz \& Canel (1996) have also shown that the
$<V/V_{max}>$ values are different; the smaller value for longer
GRBs suggests that they are on average at bigger cosmological distances.
Keeping all this in mind it seems to be definite
that these two types are physically different
objects at different cosmological scales.

Contrary to this, we did not find any significant difference in the angular
distribution of HE and NHE subclasses. This is an unclear result, because
in Pendleton et al. (1997) it is clearly stated that the $<V/V_{max}>$
values are different for the HE and NHE subclasses, and hence they
should also be at different distances. The isotropy tests do not confirm
this expectation. This also means that the separation
based on the most energetic channel remains unclear. In fact, the question
of fourth channel is highly topical recently, because
Bagoly et al. (1998) shows - independently on Pendleton et al. (1997) -
that $f4$ alone is an important quantity.
The question of most energetic channel trivially needs further study,
and is planned to be done.

  At the end of Section 3 we pointed out that the dependence
of the BATSE detection probability on the declination may mimic some sort of
anisotropy. Therefore, without a detailed correction
in accordance with the sky exposure function one
may state only the presence of an intrinsic anisotropy from the different
behaviour of short and long bursts. The requirement of the study of
this correction for both subclasses is a triviality, and is planned to
be done in the near future.

One may also speculate that the short GRBs can arise in the Galaxy
and the long ones at cosmological distances. (About the cosmological origin
of long GRBs there seems to exist no doubt; see, cf., M\'esz\'aros A.
et al. 1996).
The existence of non-zero $\omega_{2,-1}$ and probably also of
$\omega_{2,-2}$ quadrupole
terms with the simultaneous zeros for other dipole and quadrupole terms
is a strange behavior for any objects arising in the Galaxy.
E.g., it is highly
complicated to have an $\omega_{2,-1}$ term, and simultaneously not to have
the $\omega_{2,0}$ term, if the sources have arisen in the Galaxy.
 Simply, any objects in the Galaxy should have fully different
anisotropy terms (for further details and for the survey of earlier
studies of isotropy see,
e.g., Briggs 1993, 1995; Briggs et al. 1996; Tegmark et al. 1996a,b;
Meegan et al. 1996).

Remark also the following. The so called transition
scale to homogeneity (cf. M\'esz\'aros A. 1997) is minimally of size
$\simeq 300 h^{-1}$ Mpc ($h$ is the Hubble constant in units
100 km/(s Mpc)). This means that up to this distance an
inhomogeneous and anisotropic spatial distribution is not only possible but
it is even expected. In
addition, at the last time several observational implications both from the
distribution of galaxies and from the anisotropy of cosmic microwave
background radiation highly
query the fulfilment of homogeneity and isotropy even up to the
 the Hubble scale
(Lauer \& Postman 1994; \v Slechta \& M\'esz\'aros A. 1997; M\'esz\'aros A. \&
Van\'ysek 1997; Coles 1998; Sylos-Labini et al. 1998). Hence, as anisotropy
concernes we think that the different distribution of short and long GRBs
and their
cosmological origin are not in contradiction, and we further mean that
all GRBs are at cosmological distances.

\section{Conclusion}

The results of paper may be collected as follows:

A. The distribution of 2025 GRBs is anisotropic.

B. This anisotropy is not caused exclusively by instrumental effects.

C. Separating GRBs into the short and long subclasses it is shown that
the short ones are distributed anisotropically, but the long ones can still
be distributed isotropically.

D. No such difference was found for the HE-NHE separation.

E. We conjectured that the anisotropic distribution of short GRBs
does not query their cosmological origin.

\begin{acknowledgements}

Thanks are due for the valuable discussions with Z. Bagoly, V. Karas, H.M. Lee,
P. M\'esz\'aros, and an anonymous referee. One of us (A.M.) thanks
for the kind hospitality at ESO. This paper was partly supported by
OTKA grant T 024027 (L.G.B.), by GAUK grant 36/97 and
by GA\v{C}R grant 202/98/0522 (A.M), and by KOSEF grant (I.H.).

\end{acknowledgements}

\end{document}